\documentclass[aps,prl,twocolumn,superscriptaddress,showpacs]{revtex4}
\usepackage{amsmath,amssymb}
\usepackage{graphicx}
\usepackage{color}

\newcommand{\dg}{\dot\gamma}

\newcommand{\deltav}{\delta\vec v}
\newcommand{\etal}{{\it et al.}}

\newcommand{\avg}[1]{\langle #1 \rangle}

\begin{document}

\title{Shear thickening in granular suspensions: inter-particle friction and
  dynamically correlated clusters}

\author{Claus Heussinger}
\affiliation{Institute for Theoretical Physics,
  Georg-August University of G\"ottingen, Friedrich-Hund Platz 1, 37077
  G\"ottingen, Germany}
\begin{abstract}
  We consider the shear rheology of concentrated suspensions of non-Brownian
  frictional particles.  The key result of our study is the emergence of a
  pronounced shear-thickening regime, where frictionless particles would
  normally undergo shear-thinning.  We can clarify that shear thickening in our
  simulations is due to enhanced energy dissipation via frictional
  inter-particle forces. Moreover, we evidence the formation of dynamically
  correlated particle-clusters of size $\xi$, which contribute to shear
  thickening via an increase in \emph{viscous} dissipation.  A scaling argument
  gives for the {associated viscosity $\eta_v\sim \xi^2$}, which is in very
  good agreement with the data.
\end{abstract}




\pacs{83.80.Hj,83.60.Fg,66.20.Cy} \date{\today}

\maketitle

Concentrated suspensions of colloidal particles display interesting
non-Newtonian rheological behavior~\cite{mewis_wagnerBook2011}.  Shear
thickening, i.e. the increase of viscosity with shear rate, is among the most
well known effects, and has been studied for many years. In recent years a
picture of shear thickening has
emerged~\cite{wagner09:_shear,FLM:390973,Cheng02092011}, that is based on the
notion of hydro-clusters, long-lived particle clusters that are stabilized via
singular lubrication forces. With confocal imaging techniques it is now possible
to visualize these clusters~\cite{Cheng02092011}, and a quantitative
understanding of the connection between cluster formation and shear thickening
is within reach.

Another mechanism for shear thickening in dense non-Brownian granular
suspensions has recently been discussed in a series of
articles~\cite{PhysRevLett.103.086001,brown12JRheol,fall12:_shear,PhysRevLett.100.018301}.
The idea is that granular systems dilate, i.e. they want to expand when made to
flow. Under conditions of constant volume this leads to an increase in normal
stress and, subsequently, an increase in shear resistance. With hydrodynamic
thickening leading to a modest viscosity increase, dilation is a huge effect and
may effectively jam the suspension into a dynamically arrested
state~\cite{PhysRevLett.95.268302,PhysRevLett.90.178301}.

Here, we use computer simulations to study the role of inter-particle friction
in the shear rheology of dense non-Brownian suspensions.  Introducing a particle
stiffness $k$, it is possible to study the transition from the fluid to the
plastic flow regime (with a yield-stress $\sigma_y\sim k$) by increasing the
volume fraction $\phi$ through the jamming transition at $\phi_c$. Several
studies are concerned with frictionless particles and scaling laws have been
proposed that characterize the jamming
transition~\cite{olssonPRL2007,PhysRevLett.105.088303,heussingerPRL2009,otsukiPRE2009,Lerner27032012}.
The main result is that dense frictionless systems generically are shear
thinning~\cite{olssonPRL2007,otsukiPRE2009}. The role of friction has also been
studied in a variety of
contexts~\cite{cruz2005PRE,PhysRevLett.109.118305,PhysRevE.83.051301,PhysRevE.84.041308,PhysRevE.85.021305}
and the most important effect seems to be the mere shift of the critical density
to lower values. The exception being the work of Otsuki
\etal~\cite{PhysRevE.83.051301}, where a discontinuous jump between coexisting
fluid and solid branches has been observed. This constitutes the first example
of discontinuous shear thickening in a dry \emph{granular powder}.

In the present work on \emph{ granular suspensions}, we will recover this
discontinuity.  What is more intriguing, however, is a second regime of
``continuous'' shear thickening, which we explain from the enhanced viscous
dissipation of dynamically correlated particle clusters.


{\it Model~--~} We consider a two-dimensional ($d=2$) system of $N$ soft spherical
particles.  The particle volume (area) fraction is defined as $\phi =
\sum_{i=1}^N \pi R_i^2/L^2$, where $L$ is the size of the simulation box and
$R_i$ is the radius of particle $i$. To avoid crystallization, we take one half
of the particles (``small'') with radius $R_s=0.5d$
, the other half (``large'') with radius $R_l=0.7d$. Periodic (Lees-Edwards)
boundary conditions are used in both directions.

Particles interact via a standard spring-dashpot interaction (similar to
e.g.~\cite{PhysRevE.65.031304,PhysRevE.83.051301,PhysRevE.84.041308}).  Two
particles $i,j$ interact when they are in contact, i.e. when their mutual
distance $r$ is smaller than the sum of their radii $R_i+R_j$.  The normal
component of the interaction force is $F_n=k_n(r-(R_i+R_j)) -\gamma_n \delta
v_n$, where $k_n$ is the spring constant, $\gamma_n$ the dashpot strength and
$\delta v_n$ the relative normal velocity of the two contacting particles.  The
tangential component is $F_t=k_t\delta_t$, with $\delta_t$ the tangential
(shear) displacement since the formation of the contact. The tangential spring
mimics sticking of the two particles due to dry friction.  These frictional
forces are limited by the Coulomb condition $F_t\leq \mu F_n$, {with a
  constant, i.e. velocity independent friction coefficient $\mu$.}


The system is sheared at a shear rate $\dg$. Newton's equations of motion
$m\vec{\ddot r}_i = \vec F_i^{\rm cont}+\vec F_i^{\rm visc}$ are integrated with
contact forces as specified above and a viscous drag force, which implements the
shear flow. The drag force $ {\vec F}^{\rm visc}({\vec v}_i) = - \zeta
\deltav_i$, is proportional to the velocity difference $\deltav_i=\vec v_i-\vec
v_{\rm flow}$ between the particle velocity ${\vec v}_i$ and the flow velocity
${\vec v_{\rm flow}(\vec r_i)}=\vec e_x \dg
y$~\cite{durianPRL1995,scalaJCP2007,olssonPRL2007,landerEPL2010}.  The friction
coefficient $\zeta$ represents the viscosity of the surrounding fluid,
$\zeta\propto \eta_f$. Fluctuations of the flow field as well as hydrodynamic
interactions, in particular lubrication forces, are neglected. {Note that
  this automatically excludes hydrodynamic forces as possible origin for the
  shear thickening phenomena that will be discussed below. In fact, this
  tailoring of the interaction forces is a key ingredient of our study, because
  it allows to pin-point the ultimate cause of the shear thickening in the
  frictional component.}

As units we choose particle mass density $\rho$, particle diameter $d$ and the
spring constant $k_n$.  With these definitions we perform molecular dynamics
simulations using LAMMPS~\cite{lammps} with parameters $\gamma_n=0.1$,
$k_t=2k_n/7$, a static friction coefficient $\mu=1$, viscous drag $\zeta=0.1$
and a time-step of $\Delta t=0.01$. System sizes range from $N=2500$ to $4900$
particles, with a few simulations ranging up to $N=10000$.

The limit $\mu\to 0$ corresponds to the frictionless scenario, which has been
studied, for example, in
\cite{olssonPRL2007,otsukiPRE2009,PhysRevLett.105.088303}. In these systems
jamming is associated with shear-thinning rheology, governed by a critical point
at $\phi_c\approx 0.843$ and at \emph{zero stress}, $\sigma_c=0$.  {We will
  show in the following how a simple change to finite and constant friction
  coefficient $\mu\neq 0$ can fundamentally change this picture.}

\begin{figure}[ht]
  \centering
    \includegraphics[width=0.5\textwidth]{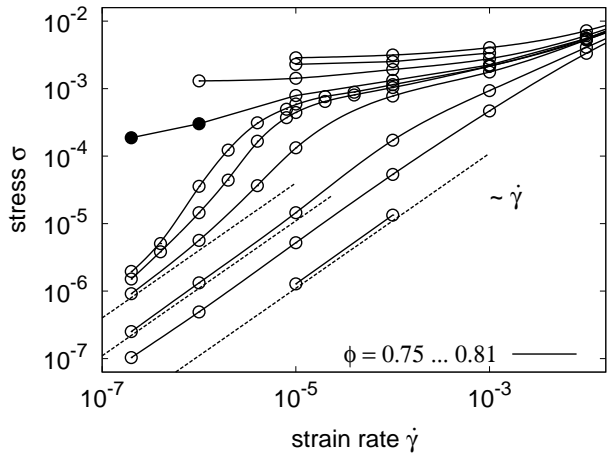}
    \caption{Flowcurves $\sigma(\dot\gamma)$ for various volume-fractions
      $\phi=0.75,0.77,0.78,0.79,0.7925,0.7935,0.795,0.8,0.805,0.81$ (from bottom to
      top).}\label{fig:flowcurve}
\end{figure}

\begin{figure}[ht]
  \centering
  \includegraphics[width=0.5\textwidth]{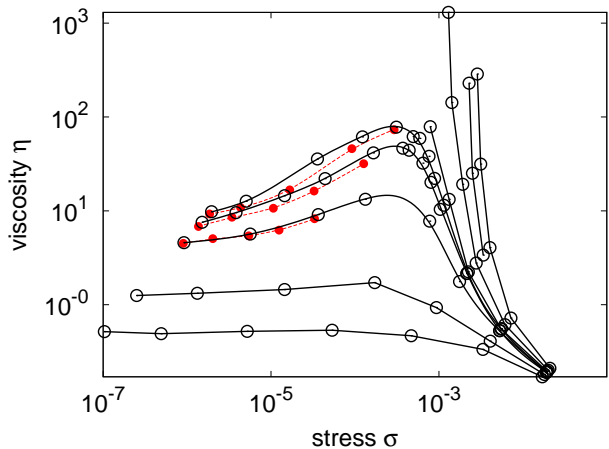}
  \caption{(Color online) Viscosity $\eta=\sigma/\dot\gamma$ vs. stress $\sigma$
    for various volume-fractions $\phi=0.77\ldots 0.81$ ($N=4900$). {As a
      comparison the data from the $N=10000$ system are given with small (red)
      symbols.}  }\label{fig:viscosity}
\end{figure}

{\it Results~--~} In Figs.\ref{fig:flowcurve} and \ref{fig:viscosity} we display
the flowcurves and the associated viscosities of our frictional simulations. By
varying the volume fraction we go through the jamming transition and observe the
associated changes in the flow behavior. At small volume fractions, below the
jamming transition, we observe a Newtonian regime $\sigma =\eta_0 \dg$, with a
strainrate-independent viscosity $\eta_0(\phi)$ that increases with volume
fraction. At high densities, above jamming, the stress levels off at the yield
stress, $\sigma_y(\phi)=\sigma(\dg\to0,\phi)$.

In frictionless systems the jamming transition is associated with ``critical''
shear-thinning $\sigma\sim\dg^x$ ($x<1$, power-law
fluid)~\cite{olssonPRL2007,otsukiPRE2009,PhysRevLett.105.088303}. Here,
surprisingly the opposite is happening: jamming is signalled by a shear
thickening regime that grows stronger with increasing the volume fraction. At
$\phi=0.78$ only a mild increase of the viscosity is observed, before it drops
in the shear thinning regime. At $\phi=0.7935$ the viscosity already increases
by about an order of magnitude!

\begin{figure}[t]
  \centering
  \includegraphics[width=0.5\textwidth]{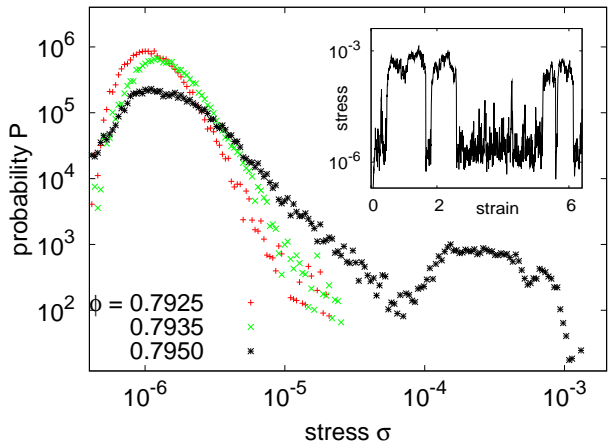}
  \caption{(Color online) Probability distribution of stress values for
    different volume fractions $\phi$ and for $\dg = 2\cdot 10^{-7}$.  The
    double-peak structure (for $\phi=0.795$) indicates the coexistence of jammed
    and viscous flow regimes. (inset) Stress-strain relation in the coexisting
    state.}\label{fig:coexistence}
\end{figure}

The stress-scale in the thickening regime (as characterized, for example, by the
stress at the viscosity maximum) is nearly independent of volume fraction.  By
way of contrast, the strainrate for the onset of thickening decreases with
volume fraction (thickening regime shifts to the left in
Fig.~\ref{fig:flowcurve}). This shift does not go down to $\dg\to 0$. Rather, at
about $\phi=0.795$, the filled data points in Fig.\ref{fig:flowcurve} indicate
qualitatively new behavior: the \emph{coexistence} of jammed solid and freely
flowing fluid states. This is evidenced in Fig.\ref{fig:coexistence}. For the
filled data points the stress distribution is bimodal (black star) and the
stress-strain relation shows sudden switching events from low-stress (fluid) to
high-stress (solid) states. By way of contrast, in the (continuous) thickening
regime (red plus, green cross) the stress distributions have only one peak. As
can be seen in the figure, the tails of this distribution are rather broad
indicative of giant stress fluctuations.

{\it Discussion~--~} The observed phenomena are strongly reminiscent of critical
behavior as described, e.g. by the van-der-Waals equation of state. The
coexistence of flowing and jammed states then signals a discontinuous jamming
transition (similar to the dry granular flow of Ref.~\cite{PhysRevE.83.051301}).
The coexistence region seems to be terminated by a ``critical point'' at a
certain (non-zero) value of stress, an associated strainrate and a volume
fraction $(\sigma_c,\dg_c,\phi_c)$, at which the transition is continuous. The
shear-thickening regime then corresponds to the near-critical ``isotherms''
close to but above this point.

Evidence of this scenario of a finite-stress critical point is provided by the
fact that stress fluctuations in the shear thickening regime are strongly
enhanced. Equally important, a large correlation length indicates cooperative
behavior. To extract such a lengthscale we calculate the velocity correlation
function $C_v(x)=\langle v_y(x)v_y(0) \rangle$, where we concentrate on the
velocity component in gradient direction, $v_y$, of two particles separated by
$x$ in the flow direction. In the frictionless system this correlation function
has been used to evidence a correlation length that diverges in the limits
$\phi\to\phi_c\approx 0.843$ and $\sigma\to
\sigma_c\equiv0$~\cite{heussingerPRL2009,olssonPRL2007}.

\begin{figure}[h!]
  \centering
  \includegraphics[width=0.5\textwidth]{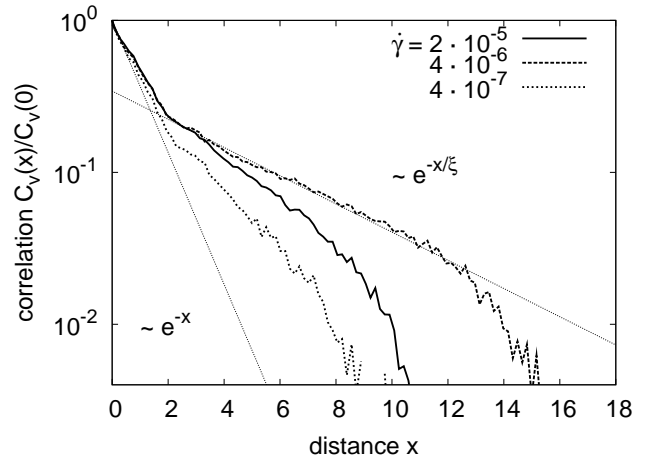}
  \caption{Velocity correlation function  $C_v(x)=\avg{{v}_y(x) v_y(0)}$ for
    different strainrates and $\phi=0.7935$}.\label{fig:velocorr}
\end{figure}

Fig.~\ref{fig:velocorr} displays the normalized correlation function for
$\phi=0.7935$ and a selected set of strainrates. Beyond a short-range
exponential decay, $C_v(x)\sim \exp(-x)$, there is clear non-monotonic behavior
with strainrate $\dg$, indicating a maximal correlation range at some finite
value $\dg_c$.
This observation can be quantified by defining the lengthscale $\xi$ from
fitting a second exponential, $C_v \sim \exp(-x/\xi)$, as indicated in the
figure~\footnote{We have checked that alternative definitions for the
  lengthscale $\xi$ do not change the resulting picture.}.

\begin{figure}[h!]
  \centering
  \includegraphics[width=0.5\textwidth]{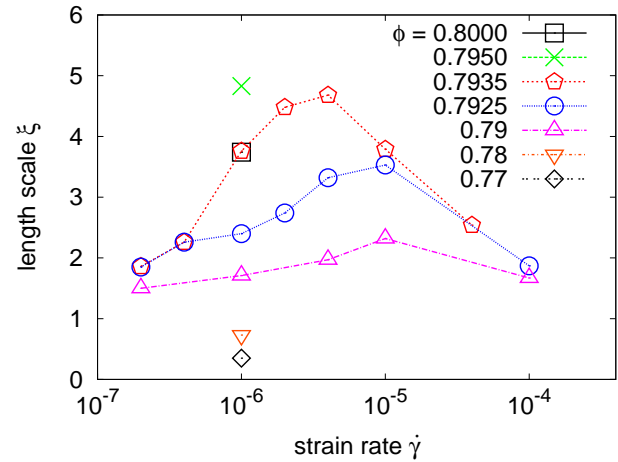}
  \caption{(Color online) Correlation length as extracted from the exponential
    fit to $C_v(\xi)$ for different volume fractions and
    strainrates.}\label{fig:lengthscale}
\end{figure}

The resulting correlation length is displayed in Fig.~\ref{fig:lengthscale}. It
clearly shows non-monotonic behavior both in strainrate $\dg$ and in
volume-fraction $\phi$. The position of the absolute maximum is estimated to be
at $\phi_c\approx 0.795,\dg_c\approx 2\cdot10^{-6},\sigma_c\approx 10^{-4}$,
which may serve as a first proxy to the critical point (see below for further
discussion).

Note, that in the frictionlesss scenario of Ref.~\cite{olssonPRL2007} the
correlation length is defined from the minimum of $C_v(x)$. We also observe a
minimum, and its behavior is similar to the $\xi$ we define. However,
finite-size effects due to the periodic boundary conditions are much stronger
and prohibit a quantitative evaluation.

{\it Relation to experiment~--~} The phenomenology described here is remarkably
similar to the experiments of Lootens
\etal~\cite{PhysRevLett.95.268302,PhysRevLett.90.178301} as well as those of
Brown \etal~\cite{brown12JRheol} and Fall \etal~\cite{fall12:_shear}. As in the
experiments we observe giant stress fluctuations in the thickening regime, as
well as coexistence of flowing and jammed states. Moreover, like in the
experiments the normal stress $p$ is tightly coupled to the shear stress
$\sigma$, such that the effective friction coefficient $\mu=\sigma/p$ is
constant ($\approx 0.3$) throughout the thickening regime (not shown).  Thus, it
seems that dilatancy effects are at the origin of the shear thickening regime.

Unlike the experiments of Brown and Fall, however, we do not observe shear
localisation. Our system is homogeneous and the flow profile is linear.
Furthermore, a tight coupling between shear and normal stresses is also observed
in simulations of frictionless particles, with either Newtonian or even
shear-thinning
behavior~\cite{heussingerSoftMatter2010,olssonPRE2011scaling.corrections,peyneauPRE2008}.
Therefore, beyond enhanced normal stresses one has to allow for a new channel of
energy dissipation via frictional particle interactions.  Such a channel is
absent in frictionless systems.

\begin{figure}[h!]
  \centering
  \includegraphics[width=0.5\textwidth]{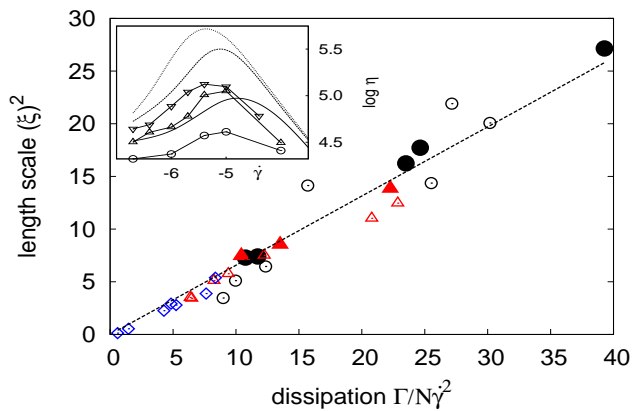}
  \caption{(Color online) Inset: Comparison of viscosity (logarithmic y-axis) as taken from
    Fig.\ref{fig:flowcurve} (thin lines) and as determined from the viscous
    dissipation $\Gamma/\dg^2$ (symbols).  Main panel: scatter plot of viscous
    dissipation $\Gamma/N\dg^2$ vs.  correlation length $\xi^2$ (circles
    $\phi=0.7935$, triangles $0.7925$, diamonds $0.770\ldots 0.790$; small open
    symbols $N=4900$, large closed symbols $N=6400\ldots 8100$).  There is a
    clear linear relation, indicating $\Gamma \propto
    N\dg^2\xi(\dg)^2$}.\label{fig:lengthscale.diss.energy}
\end{figure}

In Fig.~\ref{fig:lengthscale.diss.energy} (inset) we compare the work performed
by the external forces ($W=L^2\eta\dg^2$) with the energy dissipated by the
viscous forces ($\Gamma = -\zeta N\langle \delta v^2\rangle$). Without friction,
both should be equal to each other, so that the difference is due to energy
dissipation via friction. We see that, indeed, the shear thickening regime
corresponds to an enhanced frictional contribution to the shear thickening.
However, and perhaps surprising, even the pure viscous forces do show some
thickening behavior.

To explain this latter contribution, we need to remember that shear thickening
in our system is tightly connected to the growth of a correlation length.  If
particles move in correlated clusters of size $\xi$, then the typical velocity
scales as $\delta v\sim \dg\xi$. This leads to a renormalized energy dissipation
$\Gamma\sim \dg^2\xi^2$ and associated viscosity $\eta_v(\dg)\sim \xi(\dg)^2$.
This relation is plotted in the main panel of
Fig.~\ref{fig:lengthscale.diss.energy}. It holds remarkably well with a
prefactor of order unity.

A similar argument holds in frictionless
systems~\cite{PhysRevLett.109.105901,heussingerEPL2010}, where the relation
between correlation length, and velocity fluctuations can be used to rationalize
the divergence of the (Newtonian) viscosity with increasing the volume fraction
towards the close packing limit, $\eta(\phi)\sim \xi(\phi)^2$. In this picture,
the viscosity diverges at close packing because of the growth of dynamically
correlated particle clusters and an associated divergence of velocity
fluctuations~\cite{PhysRevLett.109.105901,heussingerEPL2010}.

With the equivalence between correlation length and viscosity $\eta_v$, we have
to reconsider the nature and location of the critical point. A divergence of the
correlation length should equally be visible as divergence in the viscosity.
However, as discussed in Ref.~\cite{brown12JRheol}, the shear thickening regime
is limited from above by an appropriate energy scale which represents the
softest link in the system (there, surface tension of the air-fluid interface).
The viscosity can therefore not grow beyond this scale. In our system this
energy scale is played by the stiffness $k_n$ of the particles. When the
viscosity $\eta_v\sim \zeta\xi^2$ of the thickening fluid is comparable to the
yield-stress $\sigma\sim k_n$ in the plastic flow regime, then thickening stops.
For the critical point, this means that it may be hidden within the plastic flow
regime.  Hard-sphere simulations, similar to Ref.~\cite{Lerner27032012} could
give valuable information in this regard.

In conclusion: we discuss the shear rheology of a non-Brownian suspension of
soft spherical particles. Hydrodynamic interactions are neglected and we
concentrate on the effects of frictional particle interactions,{
  characterized by a constant friction coefficient $\mu$. This tailoring of the
  interaction forces is a key advantage of our study.}  With this we can show
that friction does indeed lead to pronounced shear thickening, unlike in
frictionless systems which are shear-thinning.  Friction is therefore an
essential ingredient for the thickening behavior observed. {Note, that
  similar shear thickening phenomena with more complex interaction forces have
  been presented just recently in Refs.\cite{PhysRevLett.111.108301,denn} .
  Going beyond these studies we observe}
giant stress fluctuations and a growing correlation length, which is maximal
deep within the thickening regime. We show that thickening is partly due to
enhanced energy dissipation via frictional interactions. In addition,
dynamically correlated clusters of size $\xi$ also lead to an increased
\emph{viscous} contribution to the energy dissipation.  A scaling argument gives
{for the associated viscosity} $\eta_v\sim \eta_f\xi^2$, which is in very
good agreement with the data.

\acknowledgments Fruitful discussions with M. Grob and A. Zippelius are
acknowledged. Financial support comes from the Deutsche Forschungsgemeinschaft,
Emmy Noether program: He 6322/1-1.


\end{document}